\documentclass[11pt,margin=0.5in]{article}
\usepackage[utf8]{inputenc}
\usepackage[T1]{fontenc}
\usepackage{lmodern}
\usepackage{hyperref}
\usepackage{graphicx}
\usepackage{subfig}
\usepackage{comment}
\usepackage{blindtext}
\usepackage{geometry}
\usepackage{bm}
\usepackage{xcolor}
\usepackage{amssymb, amsmath, amsfonts, amsthm, mathtools}
\usepackage{lineno}

\usepackage[compress]{cite}

\title{Input Driven Synchronization of Chaotic Neural Networks with Analyticaly Determined Conditional Lyapunov Exponents}

\author{\normalsize{Jordan Culp$^{1,2,3}$ and Wilten Nicola$^{1,2,3,*}$}\\
\normalsize{$^1$Department of Physics $\&$ Astronomy, University of Calgary}\\
\normalsize{$^2$Department of Cell Biology and Anatomy, University of Calgary}\\
\normalsize{$^3$Hotchkiss Brain Institute, University of Calgary}\\
\normalsize{$^*$ Corresponding author: wilten.nicola@ucalgary.ca}
}
\date{\today}

\begin{document}
\maketitle
\begin{abstract}
Recurrent neural networks (RNNs) with random, but sufficiently strong and balanced coupling display a well known high-dimensional chaotic dynamics.  Here, we investigate if externally applied inputs to these RNNs can stabilize globally synchronous, input-dependent solutions, in spite of the strong chaos-inducing coupling.  We find that when the balance between excitation and inhibition is exact, that is when the row-sum of the weights is constant and 0, a globally applied input can readily synchronize all neurons onto a synchronous solution.  The stability of the synchronous solution is analytically explored in this work with a master stability function.  For any synchronous solution to the network dynamics, the conditional Lyapunov spectrum can be readily determined, with the stability of the synchronous solution critically dependent on the largest real eigenvalue component of the RNN weight matrix.  We find that the smaller the maximum real component of the weight matrix eigenvalues, the more readily the network synchronizes.   Further, the conditional Lyapunov exponents are easily computed numerically for any synchronization signal without simulating the RNN.  Finally, for certain oscillatory synchronization signals, the conditional Lyapunov exponents can be determined analytically. 
\end{abstract}

\section{Introduction} 

Chaotic recurrent neural networks (RNNs) have proven to be a powerful tool in analyzing the dynamics and function of populations of biological neurons \cite{rnn1,rnn2,rnn3,rnn4} , and as a general tool in machine learning to learn the dynamics behind signals and sequences \cite{force1,force2,kr1,maria1,maria2,joern1,force3}.    These networks are often initialized in a standard ``balanced" fashion, where the excitatory connection weights on average are matched by the inhibitory connection weights, leading to a high-dimensional chaotic state \cite{rnn1,rnn2,rnn3}.   This has both served as a model for excitatory/inhibitory balance in biological circuits, as well as a useful initial configuration for networks where learning algorithms simultaneously shape the structure and function of RNNs in a goal-directed manner \cite{force1,force2,force3}

While learning can produce stable dynamics in otherwise chaotic RNNs, common inputs into the network have also been shown to suppress chaotic dynamics, but not fully synchronize the neurons \cite{kr2,cn1,sc1,sc2}.  Notably, recent work has considered the role of synchronized common inputs and independent inputs in facilitating the suppression or (lack thereof) of chaotic dynamics \cite{rainer1}.  For example, in \cite{rainer1}, the authors extend the dynamic mean-field theory (DMFT) approach to analyzing chaotic RNNs.  They show that an inhibitory (but input-balanced) network fed with common oscillatory inputs effectively tracks and cancels these inputs with the recurrent dynamics.    Independent (in phase) inputs could however destabilize the chaos \cite{rainer1}.   For oscillatory inputs, the frequency of oscillatory inputs is also a critical factor in eliminating/suppressing chaotic \cite{kr2}.  Collectively, these results imply some suppression or modulation of the chaotic dynamics possible with common inputs.  However, the ability to fully synchronize the network, with either common or independent inputs to the neurons has not been considered. 

In this work, we also consider the impact of common inputs to all neurons in an otherwise chaotic RNN.  However, we investigate the conditions under which the inputs will fully synchronize all neurons in the RNN to a common solution.   First, we show that under precise, row-balancing of the excitatory and inhibitory components on the RNN weight matrix, there always exists a synchronous solution driven by common inputs.  When the weight matrix is not row-balanced, the synchronous solution still exists however it requires independent (different) inputs for each neuron.  Further, we analytically derive the stability conditions through the master stability function \cite{msf1} on the input-induced synchronous solution.   The stability condition depends only on the real components of the eigenvalues of the RNN weight matrix, and properties of the transfer function of the neurons and the synchronous solution.  As the MSF approach was utilized, the derivation does not rely on DMFT, and only assumes a diagonalizable weight matrix.  The stability of the synchronous solution was determined by computing the full conditional Lyapunov spectrum \cite{cl1,cl2}, which is easily calculable numerically by approximating a single limit.  For certain oscillatory signals, the Lyapunov spectrum can be analytically determined.  The stability results were numerically confirmed for simulated RNNs with different weight configurations and different synchronization signals.

\section{Results}
The equations for a standard recurrent neural network \cite{rnn1,rnn2} with common inputs are given by: 
\begin{eqnarray}
\dot{x}_i = -x_i + \sum_{j=1}^N \omega_{ij} \phi(x_j) + c(t), \quad i=1,2\ldots N \label{rnn1}
\end{eqnarray}
where $c(t)$ is the externally applied input to all neurons.   This input will drive the RNN onto a particular trajectory, with the RNN acting as the response system \cite{cl1,cl2}.  The function $\phi(x)$ is the transfer function or firing rate for the neuronal dynamics.   We will assume that $f(x)$ is sigmoidal, and $0\leq \phi'(x)\leq M$, for some constant $M>0$.    The weight matrix is assumed to be random, but with precisely balanced row sums.  In particular, 
\begin{eqnarray}
\omega_{ij} = J_{ij} - \frac{1}{N} \left(\sum_{j=1}^N J_{ij}\right) \label{rsbalance}
\end{eqnarray}
where 
\begin{eqnarray*}
E(J_{ij}) = 0, \quad E(J_{ij}^2) = \frac{g^2}{N}.
\end{eqnarray*}
where $E$ is the expectation operator, and $g$ scales the coupling strength.  

The row-sum condition (\ref{rsbalance}) implies a precise balance between the excitatory ($\omega_{ij}>0$) and inhibitory ($\omega_{ij}<0$) connection weights as $\sum_{j=1}^N \omega_{ij} = 0$.  Given an input $c(t)$, we know that a synchronous solution exists with the condition (\ref{rsbalance}), by considering $x_i(t) = x_s(t)$, $i=1,2,\ldots N$.  The synchronous solution is given by the differential equation 
\begin{eqnarray*}
\dot{x_i} &=& \dot{x_s} = -x_s + \sum_{j=1}^N \omega_{ij} \phi(x_s)  +c(t) \\
&=& -x_s + c(t)
\end{eqnarray*} 
and thus the network admits a synchronous solution of:
\begin{eqnarray}
x_s(t) = \int_0^t \exp(t'-t)c(t')\,dt' + x_s(0)\exp(-t)
\end{eqnarray}

\begin{figure}[htp!]
\centering
\includegraphics[scale=0.2]{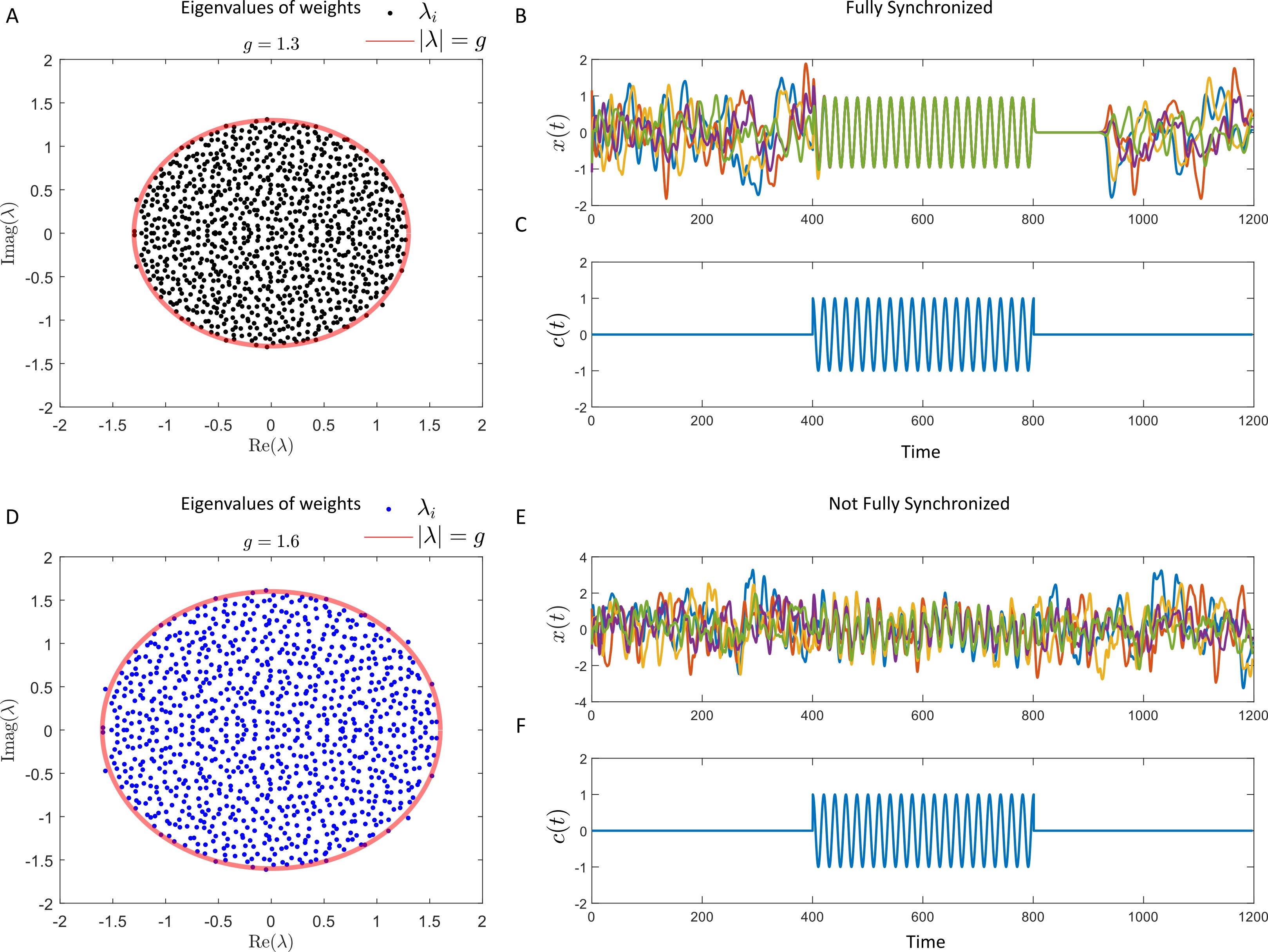}
\caption{\textbf{Full Synchronization of a Chaotic RNN} \textbf{(A)} The eigenvalues of a randomly generated weight matrix, with $g=1.3$ and $N=1000$.   The matrix is row balanced, with a constant row-sum of 0.  The red-boundary denotes the approximate eigenvalue bound $\lambda = |g|$ for a random, non-row balanced weight matrix. \textbf{(B)} The $x_i(t)$ variables for 5 neurons in a network with a common input that turns on at $t=400$.  The input is given by $c(t) = \cos(2\pi nu t)$ with $\nu =0.05$.  The input is turned on at $t=400$ and off at $t=800$.  \textbf{(C)} $c(t)$ versus time \textbf{(D)} Identical to (A), only with $g=1.6$.  Note that the randomly generated weights in (D) are rescaled (by $g$) versions of the weights in (A)   \textbf{(E)} Identical to (B), with the same $c(t)$.  The neurons no longer fully synchronize with stronger $g$. \textbf{(F)} Identical to (C). }\label{figure1}
\end{figure}

To investigate the behaviour of these networks, we simulated a network with $N=1000$ neurons with $c(t) = \cos(2\pi \nu t)$ with $\nu = 0.05$.  The synchronous solution is $x_s(t) \rightarrow \frac{2 \sin(2\pi nu t) + \cos(2\pi \nu t)}{4\pi^2 \nu^2 +1}$ for a sufficiently large $t$ (the $\exp(-t)$ decays asymptotically, Figure \ref{figure1}A-C).  For a sufficiently small $g$, but still larger than the approximate transition to chaos ($g>1$) for the unforced networks, the neurons readily synchronize to $x_s(t)$ (Figure \ref{figure1}B).   Once $c(t)$ is set to 0 after synchronization, the neurons transiently remain synchronized (although not fully) for a period of time on the system $x_i \approx x_s(t)$, with 
\begin{eqnarray}
\frac{d x_s}{dt} = - x_s \label{t0}
\end{eqnarray}
until eventually the chaotic dynamics re-emerge (Figure \ref{figure1}B-C).  For a sufficiently large $g$, the neurons no longer fully synchronize to $x_s(t)$, as $x_s(t)$ is no longer a stable synchronous solution (Figure \ref{figure1}D-F).   

In order to analyze the stability of $x_s(t)$, we will apply the Master Stability Function (MSF) approach.  Note that as we are considering the dynamics of the RNN on the synchronous solution $x_s(t)$, a DMFT analysis of the large network ($N\rightarrow \infty$) limits is unnecessary.   The main result of this work is that for any $c(t) \in C[0,\infty)$, where $C[0,\infty)$ is the set of continuous functions on $[0,\infty)$, the  conditional Lyapunov spectrum, $l_i$ of the corresponding $x_s(t)$ is determined by 
\begin{eqnarray}
l_i = -1 + \mu_i \lim_{T\rightarrow \infty} \frac{1}{T} \int_0^T \phi'(x_s(s))\,ds  = -1 + \mu_i q  ,\quad  i=1,2\ldots N\label{mr1}
\end{eqnarray} 
where $\mu_i$ is a real component of the eigenvalue of the matrix $\bm \omega$
Thus, the spectrum can be approximated without simulating the RNN in equation (\ref{rnn1}) with the following 3 step process 
\begin{itemize}
\item[1] Choose $x_s(t)$ (or $c(t)$) with $c(t)$ determined by $c = \dot{x}_s + x_s$.
\item[2] Compute $x_s(t)$ over a long-time period, $T^*$ 
\item[3] Approximate the spectrum with $l_i \approx -1 + \mu_i \frac{1}{T^*} \int_0^{T^*}  \phi'(x_x(s))\,ds =-1+\mu_i q^*$, where $ \frac{1}{T^*} \int_0^{T^*}  \phi'(x_x(s))\,ds $ is an approximation to $q$. 
\end{itemize} 
Note that the conditions $0\leq  \phi'(x)\leq M$ imply the existence of   
$$q=\lim_{T\rightarrow \infty}  \frac{1}{T} \int_0^T  \phi'(x_s(s))\,ds  $$ 
through a straightforward application of the squeeze theorem from basic calculus.

\subsection{Derivation of the Lyapunov Spectrum of $x_s(t)$}
The derivation of the Lyapunov spectrum in equation (\ref{mr1}) is routine but somewhat tedious, and follows from an immediate application of the master stability function.      First, with $x_s(t)$ determined essentially by choosing the appropriate $c(t)$ ($c(t) = x_s(t) + \dot{x}_s(t)$), we consider perturbations off of $x_s(t)$: 
\begin{eqnarray*}
\dot{x}_i = x_s(t) + \epsilon_i(t) 
\end{eqnarray*}
which yields the following:
\begin{eqnarray*}
\dot{x}_s(t) + \dot{\epsilon}_i(t)&=& -x_s - \epsilon_i +\sum_{j=1}^N \omega_{ij} \phi(x_s + \epsilon_j) + c(t) \\
\dot{x}_s(t) + \dot{\epsilon}_i(t) &=&  -x_s - \epsilon_i +\sum_{j=1}^N \omega_{ij} \left[\phi(x_s) + \phi'(x_s)\epsilon_j + O(\epsilon_j^2)\right] + c(t) \\
\end{eqnarray*}
Thus, the perturbations $\epsilon_j$, $j=1,2,\ldots N$ satisfy the variational equations to leading order 
\begin{eqnarray*}
\rightarrow  \dot{\epsilon}_i(t) &=& - \epsilon_i+\phi'(x_s(t))\sum_{j=1}^N \omega_{ij}\epsilon_j 
\end{eqnarray*}
or in matrix form: 
\begin{eqnarray*}
\dot{\bm \epsilon} =\left( -\bm I_N +  \phi'(x_s(t))\bm \omega\right) \bm \epsilon 
\end{eqnarray*}
In the MSF approach \cite{msf1}, $\bm\omega$ is assumed to be diagonalizable: 
\begin{eqnarray*}
\bm \omega = \bm P^{-1} \bm D \bm P.
\end{eqnarray*}
The substitution $\bm \eta = \bm P^{-1} \bm \epsilon$.   This yields:
\begin{eqnarray*}
\dot{\bm \eta}= \left[-\bm I_N + \phi'(x_s(t)) \bm D \right]\bm \eta  
\end{eqnarray*}
which leads to the following master-stability-function blocks:
\begin{eqnarray}
\dot{\eta} = (-1 + \lambda_i \phi'(x_s(t)))\eta \label{msf0}
\end{eqnarray} 
where $\lambda_i$ is an eigenvalue of the matrix $\omega$.   When $\lambda_i$ is a real eigenvalue, equation (\ref{msf0}) governs the dynamics of the perturbations. The solution to equation (\ref{msf0}) is determined by 
\begin{eqnarray}
\eta(t) = \exp\left(-t + \lambda_i \int_0^t \phi'(x_s(t'))\,dt'\right)\eta(0) \label{msfs1}
\end{eqnarray}

In the case of an eigenvalue having a complex conjugate pair, $\lambda = \mu_i \pm i \kappa$, then we have the following:
\begin{eqnarray*}
\dot{\eta}_+ &=& (-1 + (\mu + i \kappa)\phi'(x_s(t)))\eta_+ \\
\dot{\eta}_- &=& (-1 + (\mu - i \kappa)\phi'(x_s(t)))\eta_- 
\end{eqnarray*} 
Consider the following system:
\begin{eqnarray}
\eta_1 = \eta_+ + \eta_-, \quad \eta_2 = \frac{\eta_+ - \eta_-}{i}  
\end{eqnarray}
Then we have
\begin{eqnarray*}
\dot{\eta}_1&=& (-1 + (\mu + i \kappa)\phi'(x_s(t)))\eta_+ + (-1 + (\mu - i \kappa)\phi'(x_s(t)))\eta_- \\
&=& (-1  + \mu \phi'(x_s(t)))\eta_1 -\kappa  \phi'(x_s(t)) \eta_2\\
\dot{\eta}_2 &=& \left[ (-1 + (\mu + i \kappa)\phi'(x_s(t)))\eta_+ - (-1 + (\mu - i \kappa)\phi'(x_s(t)))\eta_- \right]i^{-1}\\
&=& (-1+\mu(f'(x_s(t)))\left(\eta_+ - \eta_- \right)i^{-1} + \phi'(x_s(t))\kappa \phi'(x_s(t))(\eta_+ +\eta_-)\\
&=& (-1+\mu(\phi'(x_s(t)))\eta_2+ \phi'(x_s(t))\kappa \eta_1
\end{eqnarray*}
And thus we have the block system $\bm \eta=(\eta_1,\eta_2)$
\begin{eqnarray}
\dot{\bm \eta} = \begin{pmatrix} -1+\mu_i\phi'(x_s(t)) & -\kappa_i \phi'(x_s(t))\\ \kappa_i \phi'(x_s(t)) & -1+\mu_i \phi'(x_s(t))   \end{pmatrix}  \bm \eta \label{bs2}
\end{eqnarray}  
Fortunately, equation (\ref{bs2}) has an exact solution which can be determined by using the matrix exponential.  The expanded form of the exact solution is:  
\begin{eqnarray}
\eta_1(t) &=& \exp\left(-t+\mu_i \int_0^t \phi'(x_s(t'))\,dt'\right)\left[\eta_1(0) \cos \left(\kappa_i\int_0^t \phi'(x_s(t')) \right)-\eta_2(0) \sin \left(\kappa_i\int_0^t \phi'(x_s(t')) \right)\right] \label{msfs2} \\
\eta_2(t) &=&\exp\left(-t+\mu_i \int_0^t \phi'(x_s(t'))\,dt'\right)\left[\eta_2(0) \cos \left(\kappa_i\int_0^t \phi'(x_s(t')) \right)+\eta_1(0) \sin \left(\kappa_i\int_0^t \phi'(x_s(t')) \right)\right] \label{msfs3}
\end{eqnarray}

Thus, the master stability function blocks (for the perturbations off $x_s(t)$) can be analytically solved for in this case.  When $\lambda$ is a real eigenvalue, the solution is governed by (\ref{msfs1}), while when $\lambda$ is a complex eigenvalue, the solution is governed by (\ref{msfs2})-(\ref{msfs3}).    The asymptotic behaviour of the perturbations $\bm \eta(t)$ in both cases is critically dependent on 
\begin{eqnarray*}
 A_i(t) = \exp\left(-t+\mu_i \int_0^t \phi'(x_s(t'))\,dt'\right).
\end{eqnarray*}
which determines the amplitude of the $i$th perturbation (associated with $\mu_i$).  

Each conditional Lyapunov exponent can then be determined by  
\begin{eqnarray*}
l_i &=& \lim_{t\rightarrow\infty} \frac{1}{t} \log (A_i(t))\\
&=&  -1 + \mu_i \lim_{t\rightarrow\infty} \frac{1}{t}\int_0^t \phi'(x_s(t'))\,dt'  \quad (\mu_i \neq 0). 
\end{eqnarray*}
Thus, the Lyapunov spectrum is given by
\begin{eqnarray*}
l_0 = 0,\quad  l_i = -1 + \mu_i \lim_{t\rightarrow\infty} \frac{1}{t}\int_0^t \phi'(x_s(t'))\,dt' = -1 + \mu_i q, \quad i=1,2,\ldots N-1 
\end{eqnarray*}

For the special case where $x_s(t)$ is a periodic function, with period $\nu^{-1}$, the long term average of the integral is the average area underneath $f'(x_s(t)')$ for a single period of the oscillation: 
\begin{eqnarray}
q=\lim_{T\rightarrow\infty} \frac{1}{T} \int_0^T \phi'(x_s(t')) \,dt' =\nu \int_0^{\nu^{-1}}\phi'(x_s(t'))\,dt'  \label{osc1}
\end{eqnarray}

\subsection{Non Row-Balanced Weights} 
We remark here that if the weights are non-row balanced, then the network can still be forced to synchronize with non-common inputs.   Suppose $x_s(t)$ is a synchronous solution, and define $c_i(t)$ as the non-common inputs to each neuron.  Then: 
\begin{eqnarray*}
\dot{x}_i &=& \dot{x_s} = -x_s + \sum_{j=1}^N \omega_{ij} \phi(x_s) + c_i(t)\\
\dot{x_s} &=&-x_s + \omega_i\phi(x_s) + c_i(t)\\
\rightarrow c_i(t) &=& \dot{x}_s + x_s - \Omega_i \phi(x_s) 
\end{eqnarray*}
where 
$$ \Omega_i = \sum_{j=1}^N\omega_{ij}.$$ 
A straightforward calculation shows that for $\epsilon_i(t)$, where $x_i(t) = x_s(t) +\epsilon_i(t)$
\begin{eqnarray*}
\dot{\epsilon}_i(t) &=& -\epsilon_i +\sum_{j=1}^N \omega_{ij} \phi'(x_s)\epsilon_i, \quad i=1,2,\ldots N  
\end{eqnarray*}
which implies that the conditional Lyapunov spectrum for $x_s(t)$ in the row-balanced case (with common inputs) is identical to the non-row balanced case (with neuron-specific inputs).

\subsection{Numerical Evaluation of the Lyapunov Spectrum} 

\begin{figure}[htp!]
\centering
\includegraphics[scale=0.1]{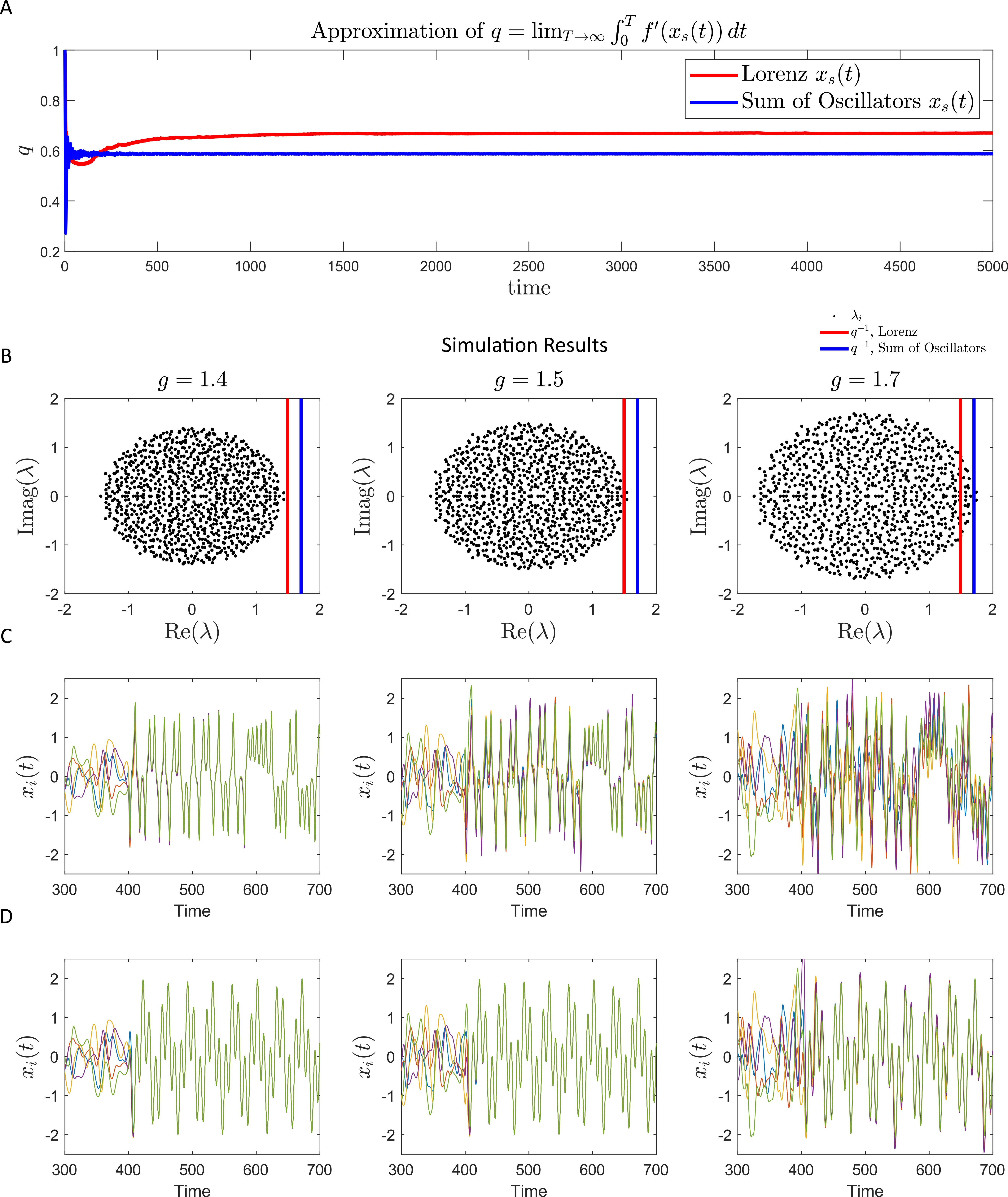}
\caption{\textbf{Numerical Results vs. Analytical Predictions} \textbf{(A)} The numerical computation of $q$.  The synchronous solution, $x_s(t)$ (Lorenz system in red, oscillator sum in blue) is simulated on $[0,5000]$, with the limit for $q$ being approximated with $q(t) = \frac{1}{t} \int_0^t f'(x_s(s))\,ds$.  The steady state value of $q(t)$ (at $t=5000$ was used) for (B). \textbf{(B)} The eigenvalues of a randomly generated, row-balanced weight matrix, as described in the main text for progressively larger $g$.  The vertical lines correspond to $q^{-1}$, the points at which $x_s(t)$ loses stability for the Lorenz solution (red) and the oscillator sum (blue).  \textbf{(C)} $x_i(t)$ for 5 neurons on the interval $[300,700]$ with $x_s(t)$ being the the $x$ component of the Lorenz system.  The initial transient from $t\in [0,300]$ is not plotted.   \textbf{(D)} Identical to (C), with the sum of oscillators $x_s(t)$. }\label{figure2}
\end{figure}

With the Lyapunov spectrum determined in equation (\ref{mr1}), the condition for the stability of $x_s(t)$ is 
\begin{eqnarray*}
\mu_i \leq q^{-1} = \lim_{T\rightarrow\infty} \frac{1}{T} \int_0^T  \phi'(x_s(s))\,ds' ,\quad  \forall \mu_i 
\end{eqnarray*}
Thus, all real-components of the eigenvalues of $\bm \omega$ must lie to the left of $q$.   To test this condition, we utilized two synchronous solutions $x_s(t)$ with synchronization signals $c(t)$ 
\begin{eqnarray*}
x_s(t)&=& \cos(2\pi t \nu_1) + \sin(2\pi t \nu_2) \quad \text{(Sum of oscillators)}  \\
x_s(t) &=& AX(t) \quad \text{(Lorenz Signal)} 
\end{eqnarray*}
where $X(t)$ is the time-rescaled $X$ component of the Lorenz system: 
\begin{eqnarray*}
\frac{dX}{dt} &=& \tau \sigma(Y-X)\\
\frac{dY}{dt} &=& \tau (X(\rho - Z) - Y)\\
\frac{dZ}{dt} &=&\tau (XY-\beta Z)
\end{eqnarray*}
The parameters for these systems can be found in Figure \ref{figure2}.   For each desired synchronous solution, the synchronization signal $c(t)$ was computed with $c(t) = \dot{x}_s(t) + x_s(t)$.  The corresponding $q$ was computed numerically in each case (Figure \ref{figure2}A).  The computed $q$ values for the two solutions were: $q\approx 0.5870$ (sum of oscillators) and $q\approx 0.6702$ (Lorenz system), which implies a loss of stability when $\mu_i > 1.4921$ (Lorenz) and $\mu_i > 1.7036$ (sum of oscillators). 

A randomly generated weight matrix, $\omega$, was created with $g$ rescaling the random weight matrix at three discrete values $1.4$, $g=1.5$ and $g =1.7$, with the synchronization signals $c(t)$ applied separately.

\subsection{An Exactly Solvable Case} 
\begin{figure}[htp!]
\centering
\includegraphics[scale=0.13]{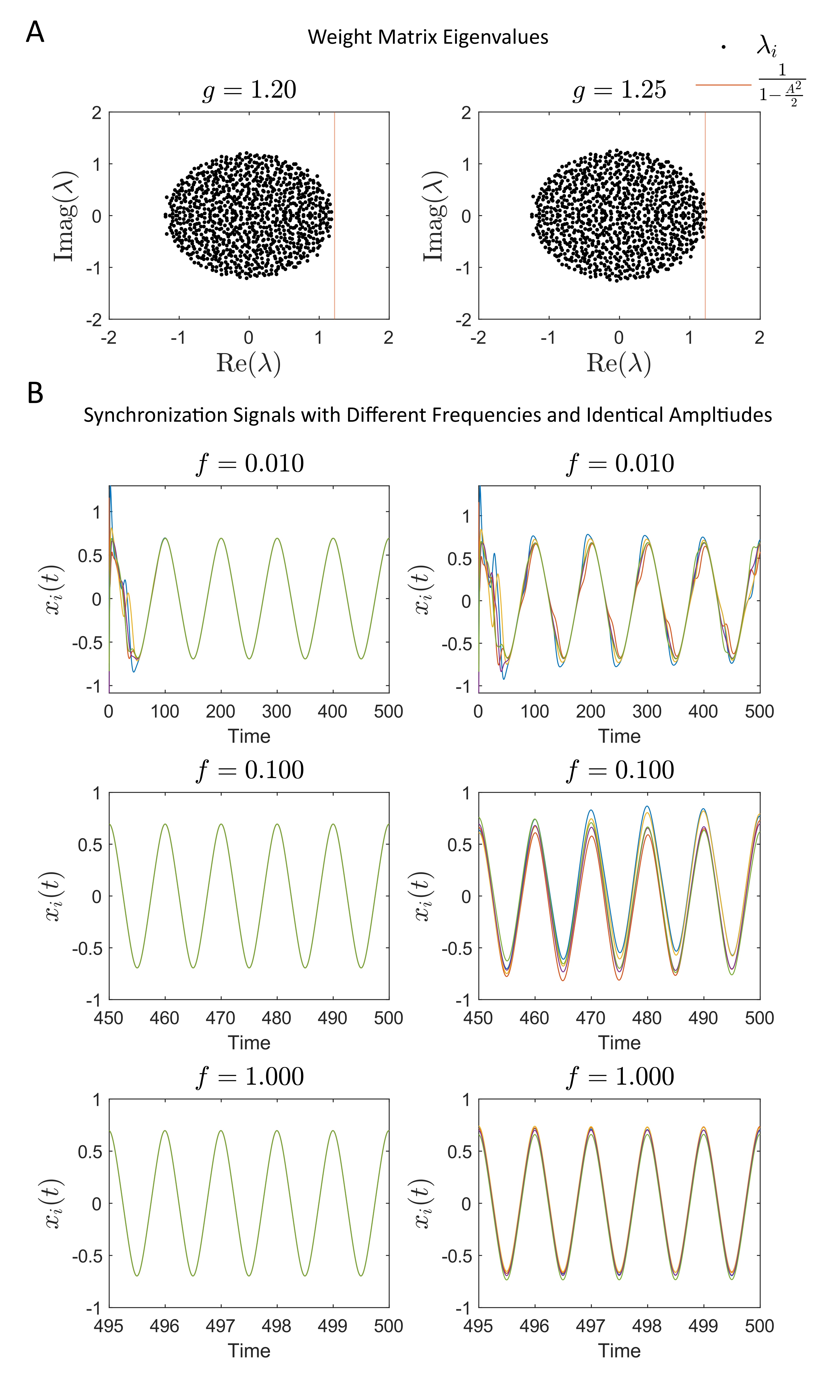}
\caption{\textbf{Frequency Independent Transition from Stable to Unstable Synchrony}  
\textbf{(A)} The eigenvalues (black dots) for a randomly generated weight matrix scaled by $g=1.2$ (left) or $g=1.25$ (right).  The vertical line corresponds to the analytically predicted loss of stability point for the synchronous solution $x_s(t) = \tanh^{-1})(\cos(2\pi f))$.  \textbf{(B)} Varying $f$ for $g=1.2$ (left) and $g=1.25$ (right).  The synchronous solution $x_s(t)$ is unstable for $g=1.25$ at all three frequency values, and stable for $g=1.2$.  The amplitude was set to $A=0.6$ in these simulations. } \label{fig3}
\end{figure}

For $f(x) = \tanh(x)$, then $f'(x) = 1-\tanh(x)^2$, the MSF blocks are given by
\begin{eqnarray*}
\dot{\eta} &=& (-1 + \lambda- \lambda\tanh(x_s(t))^2 )\eta \quad \text{(Real Eigenvalues)} \\
\dot{\bm \eta} &=& \begin{pmatrix} -1+\mu  - \mu \tanh(x_s(t))^2 & -\omega + \omega\tanh(x_s(t))^2  \\ \omega  - \omega \tanh(x_s(t))^2 & -1+\mu  - \mu \tanh(x_s(t))^2   \end{pmatrix}  \bm \eta \quad \text{(Complex Eigenvalues)} 
\end{eqnarray*}
Consider the following synchronous solution ($x_s(t)$):
\begin{eqnarray*}
x_s(t) &=& \tanh^{-1}(A\cos(2\pi f  t)), \quad A<1.  \\
c(t) &=&  \dot{x}_s + x_s \\
&=&-\frac{2A\pi f \sin(2\pi f t)}{1-A^2 \cos(2\pi f t)^2}+\tanh^{-1}(A\cos(2\pi f t))
\end{eqnarray*} 
where $f$ is the frequency of $x_s(t)$ and $A$ is the amplitude.

which yields: 
\begin{eqnarray*}
\dot{\eta} &=& (-1 + \lambda- \lambda A^2\cos^2(2\pi f t))\eta \quad \text{(Real Eigenvalues)} \\
\dot{\bm \eta} &=& \begin{pmatrix} -1+\mu  - \mu A^2\cos^2(2\pi f t) & -\omega + \omega A^2\cos^2(2\pi f t) \\ \omega  - \omega A^2\cos^2(2\pi f t) & -1+\mu  - \mu A^2\cos^2(2\pi f t)  \end{pmatrix}  \bm \eta \quad \text{(Complex Eigenvalues)} 
\end{eqnarray*}
The exponential decay of these two solutions is governed by:
\begin{eqnarray*}
-t + \mu t - \mu \int_0^t A^2 \cos^2(2\pi\nu t')\,dt' = -t + \mu t - \mu \left(\frac{A^2}{2}t +\frac{A^2}{8\pi \nu}\sin(4\pi \nu t) \right) 
\end{eqnarray*}
Thus, we require for all real $\mu$, 
\begin{eqnarray*}
-1 +\mu_i - \mu_i \frac{A^2}{2} < 0, \forall \mu_i 
\end{eqnarray*}
which implies that 
\begin{eqnarray*}
\mu_i < \frac{1}{1-\frac{A^2}{2}}, \forall \mu_i
\end{eqnarray*}
is the condition for the local asymptotic stability of $x_s(t)$.   Interestingly, the stability of the synchronous solution is not dependent in this case, on its frequency.   We tested this result over three orders of magnitude of the frequency range of $x_s(t)$ with $f=10^{-j}$ for $j=0,1,2$, and $A=0.6$. The predicted loss of stability occurs when the maximum real eigenvalue crosses the $\mu_i = 1.2195$ (Figure \ref{fig3}).  For $g=1.2$, $x_s(t)$ is stable at all three frequency values, and loses stability for $g=1.25$ (Figure \ref{fig3}A-B).   Note that a cursory evaluation of equation (\ref{osc1}) yields $q=1-\frac{A^2}{2}$.

\section{Discussion} 

The high-dimensional chaotic dynamics in RNNs has been well studied for its computational properties, and as a model for excitatory/inhibitory balance \cite{rnn1,rnn2,rnn3,rnn4,force1,force2,kr1}.   Recent work has shown that these chaotic dynamics can be partially suppressed with common inputs \cite{kr2}, although this depends on the specific network considered \cite{rainer1}.  Here, we derive the conditions at which a global and fully synchronous solution, elicited by common inputs, completely suppresses the chaotic dynamics.  In particular, if the row-sum of the chaos inducing weight matrix is exactly 0 for all neurons, a common input will synchronize all neurons, even in the chaotic regime ($g> 1$, $N\gg O(1)$, \cite{rnn1,rnn2}).  The spectrum of conditional Lyapunov exponents was determined, with a simple numerical scheme for its determination.  There is a critical point for the largest real component of the eigenvalues of the weights that causes a loss of stability in $x_s(t)$.   If any eigenvalue is larger than this crossing point, then the synchronous solution loses local asymptotic stability.   

As these results were derived with a master stability function, and do not rely on DMFT, they are globally applicable for any weight matrix coupling RNNs of the form in Equation (\ref{rnn1}).   The only constraint on the weight matrix is that it be diagonalizable.   Thus, even for low-rank perturbations of random matrices \cite{lr1} or block matrices representing differing cell types \cite{rnn3}, the stability of the synchronous solution elicited by common inputs can be readily determined if the matrices are row-balanced.  If the connection matrices are not precisely row-balanced, then independent inputs to each neuron with a common component can fully synchronize the network.

 We note that the input driven synchronization of chaotic RNNs differing from (\ref{rnn1}) has been considered by others, with a master stability function approach.   For example, smaller, time-delayed neural networks have been considered in \cite{cn1}.  The authors construct and analyze (with Lyapunov functions) the stability of a synchronous solution in recurrent neural networks coupled with time-delayed connectivity. 

Finally, we note that the analytical description of the conditional Lyapunov exponents for otherwise chaotic RNNs may be of use in validating recent advances in computing Lyapunov exponents \cite{comp1,comp2}.  However, we remark that the conditional Lyapunov exponents computed here are exclusively for the globally synchronous solution induced by an input, rather than the intrinsic chaotic dynamics considered in \cite{comp1,comp2}.

\section*{Acknowledgements }

W. N. was funded an NSERC Discovery Grant, and a Tier II Canada Research Chair in Computational Neuroscience, and through the Hotchkiss Brain Institute.  We'd like to thank Yonathan Aljadeff for his helpful suggestions in improving the quality of the manuscript.

\clearpage
\section*{Figures} 

\end{document}